\documentclass[letterpaper,final]{cjpsuppl}
\usepackage{graphicx}
\usepackage{here}
\usepackage{latexsym}
\usepackage{hyperref}

\def\url#1{{\href{#1}{\sf #1}}}
\def\urll#1#2{{\href{#1}{\sf #2}}}
\newcommand{\eprint}[1]{\mbox{\href{http://arxiv.org/abs/#1}{\sf #1}}}

\newcommand{\gev}{\hbox{ GeV}}

\newcommand{\mev}{\hbox{ MeV}}

\newcommand{\tev}{\hbox{ TeV}}

\newcommand{\cm}{\hbox{ cm}}

\def\ltap{\mathop{\raisebox{-.4ex}{\rlap{$\sim$}} 
\raisebox{.4ex}{$<$}}}
\def\gtap{\mathop{\raisebox{-.4ex}{\rlap{$\sim$}} 
\raisebox{.4ex}{$>$}}}

\newcommand{\cfrac}[2]{\textstyle \frac{#1}{#2}}

\newcommand{\m}{\hbox{ m}}

\newcommand{\onetev}{1-TeV scale}

\def\slashii#1{\setbox0=\hbox{$#1$}             
   \dimen0=\wd0                                 
   \setbox1=\hbox{\sl/} \dimen1=\wd1            
   \ifdim\dimen0>\dimen1                        
      \rlap{\hbox to \dimen0{\hfil\sl/\hfil}}   
      #1                                        
   \else                                        
      \rlap{\hbox to \dimen1{\hfil$#1$\hfil}}   
      \hbox{\sl/}                               
   \fi}                                         %
\def\slashiii#1{\setbox0=\hbox{$#1$}#1\hskip-\wd0\hbox to\wd0{\hss\sl/\/\hss}}
\catcode`\@=11
\def\ps@fnal{\def\@oddhead{\textsf{FERMILAB--CONF--05/018--T} \hfill 
\thepage}
\def\@evenhead{\thepage \hfil \textsf{FERMILAB--CONF--05/018--T}}
\let\@evenfoot\@empty
\let\@oddfoot\@empty}

\def\ps@titpg{\def\@oddhead{\textsf{\hfill FERMILAB--CONF--05/018--T}}\let\@oddfoot\@empty}

\def\ps@specialpg{\def\@evenhead{\thepage \hfil FERMILAB--CONF--05/018--T}
\let\@evenfoot\@empty}

\begin{document}
\title{Revolutions and Revelations}
\authori{Chris Quigg}
\addressi{Fermi National Accelerator Laboratory, \\ P.O. Box 500, Batavia, Illinois, 
USA 60510}
\authorii{}    \addressii{}
\authoriii{}   \addressiii{}
\authoriv{}    \addressiv{}
\authorv{}     \addressv{}
\authorvi{}    \addressvi{}
\headtitle{Revolutions and Revelations}
\headauthor{Chris Quigg}
\lastevenhead{Chris Quigg: Revolutions and Revelations}
\refnum{}
\daterec{}
\suppl{A}  \year{2004} \setcounter{page}{1}
\maketitle

\begin{abstract}
Concluding talk, \textit{Physics at LHC 2004,} Vienna
\end{abstract}

\section{Introduction}

\subsection{Vienna Tales}
It is a great pleasure to be back in Vienna, among friends, and in these splendid 
surroundings of the Academy of Sciences. Vienna is the 
birthplace of my scientific grandfather, Viki Weisskopf, and some 
years ago I had the good fortune to spend a month as Schr\"{o}dinger 
Professor in the Institute for Theoretical Physics on Boltzmanngasse.

I'd like to begin by telling you a lesson in Viennese culture that I
learned during my appointment at the University.  One fine day toward
the end of my stay, my hosts dispatched me to the bank with a very
official-looking piece of paper that I could exchange for my
Schr\"{o}dinger stipend, a stack of Schilling notes in various
denominations.  In those pre-Euro days, it seemed straightforward to
peer into a nation's cultural self-identity by examining the faces on
the banknotes.  Imagine my delight when I found Mozart on the
5000-Schilling note, and my rapture when I discovered my own
Schr\"{o}dinger on the 1000-Schilling note.  Knowing from daily
commerce that Freud was only on the 50-Schilling note, I thought to
myself, ``What a civilized country!  Austrians really must have their
priorities straight: Music!  Physics!''

I fairly sprinted back to the Institute for Theoretical Physics to
share my new cultural insight.  Now, Alfred Bartl is a very gentle man,
but when I told him what I had just learned about the Austrian soul, he
looked at me for a long moment with great pity.  Then he said, in a
patient soothing tone, ``Don't you understand \ldots no Austrian has
ever seen a Mozart or a Schr\"{o}dinger --- but Freud is everwhere!''

Perhaps I did over-interpret the faces of Mozart and Schr\"{o}dinger on
the old Schilling notes, but there is no doubt that Vienna is one of
the great sites of our patrimony as particle physicists.  Not far from
here, in the meadows of the Prater, Victor Hess launched the famous
balloon flights during which---measuring how the conductivity of the
atmosphere varies with altitude---he discovered the cosmic
radiation~\cite{federmann}. At the Radium Institute, Marietta Blau 
made enhancements to the sensitivity of photographic emulsions that 
led, in 1937, to the observation of ``stars,'' the many-body disintegration of 
nuclei under the impact of cosmic rays~\cite{blau}. 
And let us not neglect Wolfgang Pauli, whose theoretical 
insights and spooky perturbations on experimental apparatus are the 
stuff of legend~\cite{wp}. Vienna's scientific heritage is as rich as 
its musical, artistic, and literary past, and it provides an inspiring 
setting for our discussions of the future of particle physics and the 
exciting prospects for the Large Hadron Collider.

Our confidence in a vibrant future grows out of the accomplishments 
of the recent past, so let us take a moment to assess some of the 
contributions that set the scene for the LHC's era of exploration.

\subsection{A Decade of Discovery Past\label{subsec:discovery}}
We particle physicists are impatient and ambitious people, and so we tend to regard 
the decade just past as one of consolidation, as opposed to stunning 
breakthroughs. But a look at the headlines of the past ten years gives 
us a very impressive list of discoveries.
\P\ The electroweak theory has been elevated from a very 
    promising description to a \textit{law of nature.} This 
    achievement is truly the work of many hands; it has involved 
    experiments at the $Z^{0}$ pole, the study of $e^{+}e^{-}$, 
    $\bar{p}p$, and $\nu N$ interactions, and supremely precise 
    measurements such as the determination of $(g-2)_{\mu}$.
    \P\ Electroweak experiments have observed what we may 
    reasonably interpret as the influence of the Higgs boson in the 
    vacuum.
    \P\ Experiments using neutrinos generated by cosmic-ray 
    interactions in the atmosphere, by nuclear fusion in the Sun, and 
    by nuclear fission in reactors, have established neutrino flavor 
    oscillations: $\nu_{\mu} \to \nu_{\tau}$ and  $\nu_{e} \to 
    \nu_{\mu}/\nu_{\tau}$. 
    \P\ Aided by experiments on heavy quarks, studies of 
    $Z^{0}$,  investigations of high-energy $\bar{p}p$, $\nu N$, and $ep$ 
    collisions, and by developments in lattice field theory, we have 
    made remarkable strides in understanding quantum chromodynamics
    as the theory of the strong interactions.
    \P\ The top quark, a remarkable apparently elementary 
    fermion with the mass of an osmium atom, was discovered in 
    $\bar{p}p$ collisions.
    \P\ Direct $\mathcal{CP}$ violation has been observed in $K \to \pi\pi$ decay. 
    \P\ Experiments at asymmetric-energy $e^{+}e^{-} \to 
    B\bar{B}$ factories have established that $B^{0}$-meson decays do 
    not respect $\mathcal{CP}$ invariance.
    \P\ The study of type-Ia supernovae and detailed thermal 
    maps of the cosmic microwave background reveal that we live in a 
    flat universe dominated by dark matter and energy.
    \P\ A ``three-neutrino'' experiment has detected the 
    interactions of tau neutrinos.
    \P\ Many experiments, mainly those at the highest-energy 
    colliders, indicate that quarks and leptons are structureless on 
    the \onetev.

We have learned an impressive amount in ten years, and I find quite 
striking the diversity of experimental and observational approaches 
that have brought us new knowledge, as well as the richness of the 
interplay between theory and experiment. Let us turn now to the way 
the quark--lepton--gauge-symmetry revolution has taught us to view the 
world.

\subsection{How the world is made\label{subsec:how}}

\begin{figure}[t!]
\begin{center}
\includegraphics[width=6.3cm]{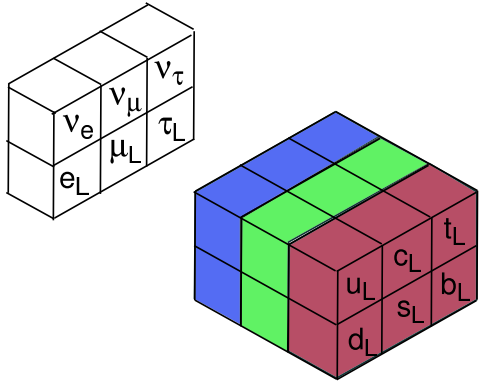}
\caption{The left-handed doublets of quarks and leptons that inspire 
the structure of the electroweak theory. \label{fig:DumbL}}
\end{center}
\vspace*{-24pt}
\end{figure}

Our picture of matter is based on the recognition of a set of pointlike 
constituents: the quarks,
\begin{equation}
\left(
		\begin{array}{c}
			u  \\
			d
		\end{array}
		 \right)_{\mathrm{L}} \;\;\;\;\;\;
		\left(
		\begin{array}{c}
			c  \\
			s
		\end{array}
		 \right)_{\mathrm{L}} \;\;\;\;\;\;
		\left(
		\begin{array}{c}
			t  \\
			b
		\end{array}
		 \right)_{\mathrm{L}}	\;,
		 \label{eq:quarks}
	\end{equation}	
and the leptons,
\begin{equation}
\left(
		\begin{array}{c}
			\nu_{e}  \\
			e^{-}
		\end{array}
		 \right)_{\mathrm{L}} \;\;\;\;\;\;
		\left(
		\begin{array}{c}
			\nu_{\mu}  \\
			\mu^{-}
		\end{array}
		 \right)_{\mathrm{L}} \;\;\;\;\;\;
		\left(
		\begin{array}{c}
			\nu_{\tau}  \\
			\tau^{-}
		\end{array}
		\right)_{\mathrm{L}}	\;,
		\label{eq:leptons}
	\end{equation}
as depicted in Figure~\ref{fig:DumbL}, 
plus a few fundamental forces derived from gauge symmetries. The quarks 
are influenced by the strong interaction, and so carry \textit{color}, 
the strong-interaction charge, whereas the leptons do not feel the 
strong interaction, and are colorless. By pointlike, we understand 
that the quarks and leptons show no evidence of internal structure at 
the current limit of our resolution,  ($r \ltap 10^{-18}\m$).

\begin{figure}[t!]
\begin{center}
\includegraphics[width=11cm]{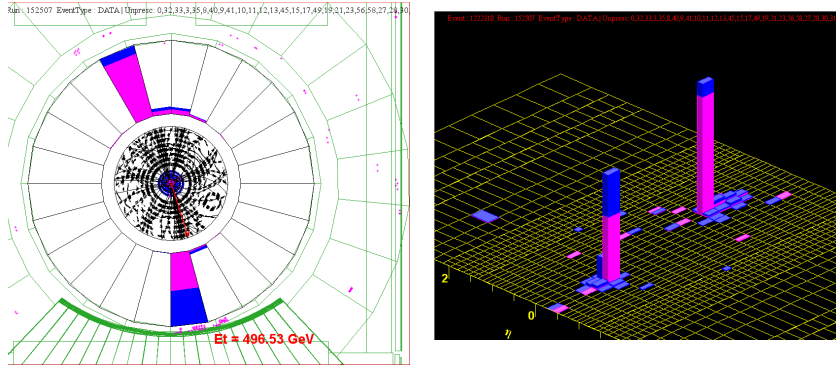}
\caption{A Tevatron Collider event with $1364\gev$ of transverse 
energy, recorded in the CDF detector. The left panel shows an end view 
of the detector, with tracking chambers at the center and calorimeter 
segments at medium and large radii. The right panel shows the 
Lego$^{\mathrm{TM}}$ 
plot of energy deposited in cells of the cylindrical detector, 
unrolled. See Ref.~\cite{Gallinaro:2003qr}.\label{fig:CDF1364}}
\end{center}
\vspace*{-12pt}
\end{figure}

The notion that the quarks and leptons are elementary---structureless 
and indi\-vis\-ible---is necessarily provisional. \textit{Elementarity} 
is one of the aspects of our picture of matter that we test ever more 
stringently as we improve the resolution with which we can examine 
the quarks and leptons. For the moment, the world's most powerful 
microscope is the Tevatron Collider at Fermilab, where collisions of 
980-GeV protons with 980-GeV antiprotons are studied in the CDF and 
D\O\ detectors. The most spectacular collision recorded so far, which 
is to say the closest look humans have ever had at anything, is the 
CDF two-jet event shown in Figure~\ref{fig:CDF1364}.
This event almost certainly corresponds to the collision of a quark 
from the proton with an antiquark from the antiproton. Remarkably, 
70\% of the energy carried into the collision by proton and 
antiproton emerges perpendicular to the incident beams. 
At a given transverse energy $E_{\perp}$, we may roughly estimate 
the resolution as $r \approx (\hbar c)/E_{\perp} \approx 2 \times 
10^{-19}\tev\m/E_{\perp}$. \footnote{See  ``Searches for Quark and 
Lepton Compositeness'' in Ref.~\cite{Eidelman:2004wy} for a more detailed 
discussion.} Imagine what the LHC will bring!

\begin{figure}[b!]
\begin{center}
\includegraphics[width=10cm]{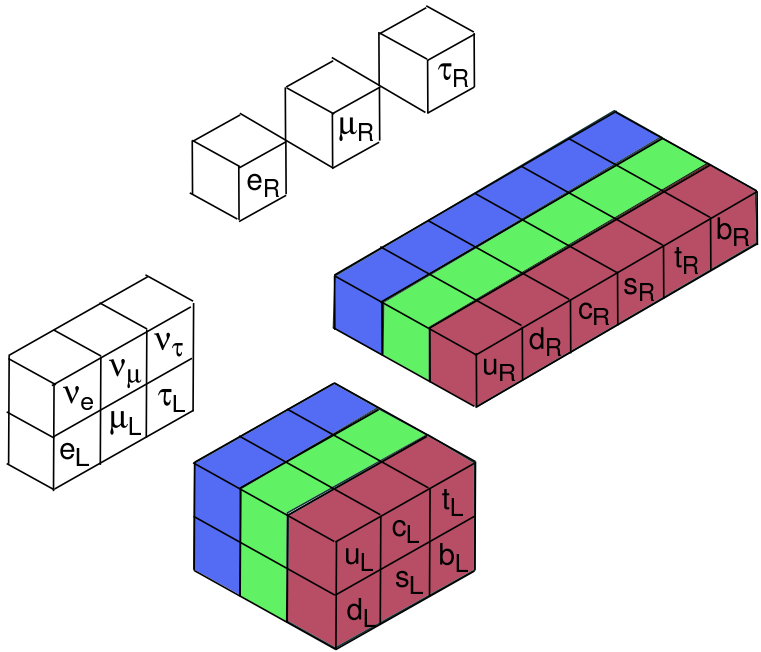}
\vspace*{-12pt}
\caption{The left-handed doublets and right-handed singlets of quarks 
and leptons.\label{fig:DumbSM}}
\end{center}
\vspace*{-24pt}
\end{figure}

Looking a little more closely at the constituents of matter, we find 
that our world is not as neat as the simple cartoon vision of 
Figure~\ref{fig:DumbL}. The left-handed and right-handed fermions 
behave very differently under the influence of the charged-current weak 
interactions. A more complete picture is given in Figure~\ref{fig:DumbSM}.
[This figure represents the way we looked at the world before the 
discovery of neutrino oscillations that require neutrino mass and 
almost surely imply the existence of right-handed neutrinos.] Neutrinos 
aside, the striking fact is the asymmetry between left-handed fermion 
doublets and right-handed fermion singlets manifested in 
\textit{parity violation} in the charged-current weak interactions. 
What does this distinction mean?

\begin{figure}[t!]
\begin{center}
\includegraphics[width=10cm]{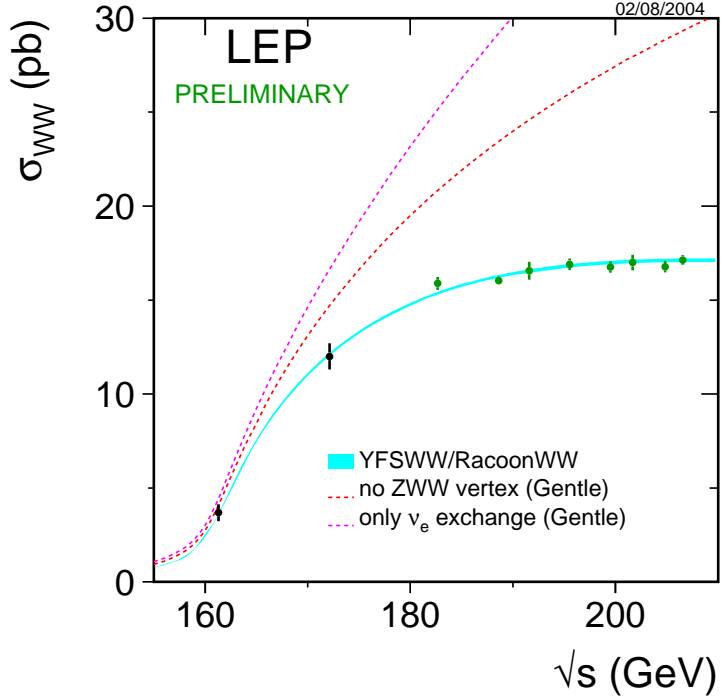}
\vspace*{-24pt}
\caption{Cross section for the reaction $e^{+}e^{-} \to W^{+}W^{-}$
measured by the four LEP experiments, together with the full
electroweak-theory simulation and the cross sections that would result
from $\nu$-exchange alone and from
$(\nu+\gamma)$-exchange~\cite{LEPEWWG}.  \label{fig:LEPgc}}
\end{center}
\vspace*{-24pt}
\end{figure}

A remarkable achievement of recent experiments is the clear 
test of the gauge symmetry, or group-theory structure, of the 
electroweak theory, in the reaction $e^{+}e^{-} \to W^{+}W^{-}$. 
Neglecting the electron mass, this reaction is described by three 
Feynman diagrams that correspond to $t$-channel neutrino exchange
and $s$-channel photon and $Z^{0}$ exchange. The LEP measurements in Figure~\ref{fig:LEPgc}
agree well with the predictions of electroweak-theory Monte Carlo
generators, which predict a benign high-energy behavior.  If the 
$Z$-exchange contribution is omitted (middle dashed line) or if both the 
$\gamma$- and $Z$-exchange contributions are omitted (upper dashed 
line), the calculated cross section grows unacceptably with 
energy---and disagrees with the measurements.  The gauge cancellation 
in the $J=1$ partial-wave amplitude is thus observed.

\begin{figure}[t!]
\begin{center}
\includegraphics[width=9cm]{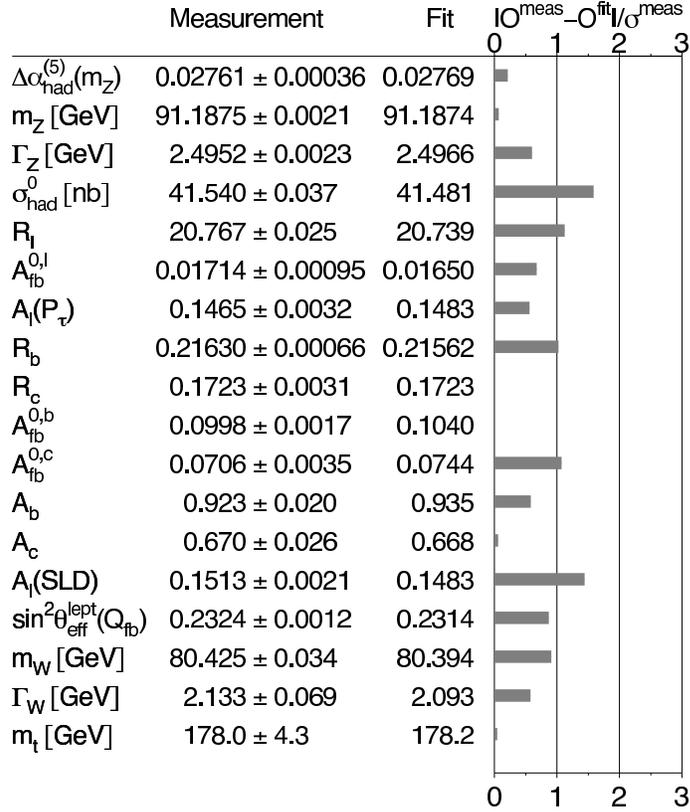}
\vspace*{-12pt}
\caption{Precision electroweak measurements and the pulls they exert 
on a global fit to the standard model, from Ref.\ 
{\protect\cite{LEPEWWG}}.}
\label{fig:pulls}
\end{center}
\vspace*{-24pt}
\end{figure}

The comparison between the electroweak theory and a considerable 
universe of data is shown in Figure~\ref{fig:pulls}
where the pull, or difference between the global fit and measured 
value in units of standard deviations, is shown for eighteen
observables~\cite{LEPEWWG}.
The distribution of pulls for this fit, due to the LEP Electroweak
Working Group, is not noticeably different from a normal distribution,
and only one measurement differs from the fit by as much as about two
standard deviations~\cite{lang}.  It is from fits of the kind represented here that
we learn that the standard-model interpretation of the data favors a
light Higgs boson.

\begin{figure}[t!]
\begin{center}
\includegraphics[width=12cm]{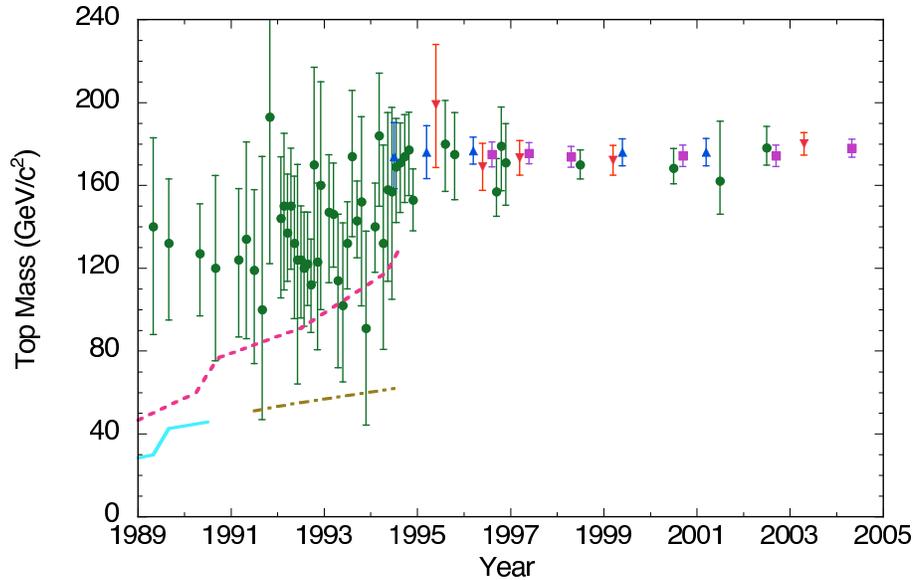}
\caption{Indirect determinations of the top-quark mass from fits to
electroweak observables (open circles) and 95\% confidence-level lower
bounds on the top-quark mass inferred from direct searches in
$e^{+}e^{-}$ annihilations (solid line) and in $\bar{p}p$ collisions,
assuming that standard decay modes dominate (broken line).  An indirect
lower bound, derived from the $W$-boson width inferred from $\bar{p}p
\rightarrow (W\hbox{ or }Z)+\hbox{ anything}$, is shown as the
dot-dashed line.  Direct measurements of $m_{t}$ by the CDF (triangles)
and D\O\ (inverted triangles) Collaborations are shown at the time of
initial evidence, discovery claim, and at the conclusion of Run 1.  The
world averages from direct observations are shown as squares.  For
sources of data, see Ref.  {\protect\cite{Eidelman:2004wy}}.  (From
Ref.\ {\protect\cite{Quigg:1997uh}}.)} \label{fig:EWtop}
\end{center}
\end{figure}

While testing the consistency of the theory, precision measurements
also give us indications of the values of unknown parameters; these
indications, in turn, set up new tests of the theoretical framework.  A
notable example is the time evolution of the top-quark mass favored by
simultaneous fits to many electroweak observables, which I show in
Figure \ref{fig:EWtop}.  Higher-order processes involving virtual top
quarks are an important element in quantum corrections to the
electroweak theory's predictions the  makes for many observables, and 
so each measurement is, in effect, an indirect measurement of the 
top-quark mass.
The success of these indirect determinations in pointing to a heavy
top-quark mass encourages us to believe that we should put some stock
in similar indications of a light Higgs-boson mass. The current accord 
between direct and indirect determinations of the top mass is shown 
in the last entry of Figure~\ref{fig:pulls}.

These are just a few indications of the quantitative successes of the
electroweak theory, which is only one component of the standard-model
edifice.  They serve as token reminders of the strong foundation on
which our hopes for future progress rest.  In what follows, I want to
very briefly point to five areas in which I believe we can anticipate
truly revolutionary progress over the next decade or two.\footnote{A
more extensive discussion is to be found in my lecture at the SLAC
Summer Institute~\cite{cqslac}.}

\section{Revolution: Understanding the Everyday}
The first revolution will be led by the LHC over the next 
decade. That is the problem of 
understanding the everyday, the stuff of the world around us. It 
pertains to basic questions:  Why are there atoms?  Why 
is there chemistry?  Why are stable structures possible?  
Knowing the answers to those questions may even give us an insight 
into What makes life possible?

Those are the general questions that we are seeking to answer when we
look for the origin of electroweak symmetry breaking.  I think that the
best way to make the connection is to consider what the world would be
like if there were nothing like the Higgs mechanism for electroweak
symmetry breaking.  First, it's clear that quarks and leptons would
remain massless, because mass terms are not permitted  if the electroweak symmetry remains
manifest.\footnote{I assume for this discussion that all the trappings
of the Higgs mechanism, including Yukawa couplings for the fermions,
are absent.} We've done nothing to the strong interaction, so QCD would
still confine the (massless) color-triplet quarks into color-singlet
hadrons, with very little change in the masses of those stable
structures.  In particular, the nucleon mass would be essentially
unchanged, but the proton would outweigh the neutron because the down
quark now does not outweigh the up quark, and that change will have its
own consequences.

Even in the absence of a Higgs mechanism, the electroweak symmetry is
broken by QCD. As we approach low energy from above, the chiral
symmetry that treated the massless left-handed and right-handed quarks
as separate objects is broken.  The resulting communication between the
left-handed and right-handed worlds engenders a breaking of the
electroweak symmetry.  It's not a satisfactory theory of our world
because the scale of electroweak symmetry breaking is measured by the
pseudoscalar decay constant of the pion, so the amount of mass acquired
by the $W$ and $Z$ is set by $f_{\pi}$, not by what we know to be the
electroweak scale: it is off by a factor of 2500.

But the fact is that QCD breaks the electroweak symmetry, so the world
without a Higgs mechanism---but with strong-coupling QCD---is a world
in which the $\mathrm{SU}(2)_{\mathrm{L}}\otimes \mathrm{U}(1)_{Y}$
becomes $\mathrm{U}(1)_{\mathrm{em}}$.  Because the weak bosons have
masses, the weak-isospin force, which we might have taken to be a
confining force in the absence of symmetry breaking, is not confining.
Beta decay is very rapid, because the gauge bosons are very light.  The
lightest nucleus is therefore one neutron; \textit{there is no hydrogen atom.}
It is likely that some light elements, such as helium, would be created
in the first minutes after the big bang.  Because the electron is
massless, the Bohr radius of the atom is infinite, so there is nothing
we would recognize as an atom, there is no chemistry as we know it,
there are no stable composite structures like the solids and liquids we
know.

Look how very different the world would be, if it were not for the
mechanism of electroweak symmetry breaking whose inner workings we
intend to explore and understand at the LHC. What we are really trying
to get at, when we look for the source of electroweak symmetry
breaking, is why we don't live in a world so different, why we live in
the world we do.  I think that's a glorious question.  It's one of the
deepest questions that human beings have ever tried to engage, and
\textit{you} will answer this question.

What could the answer be?  The agent of electroweak symmetry breaking
represents a novel fundamental interaction at an energy of a few
hundred GeV. We do not know what that force is.  It could be the Higgs
mechanism of the standard model (or a supersymmetric standard model),
which is built in analogy to the Ginzburg--Landau description of
superconductivity.  Maybe it is a new gauge force, perhaps operating on
as yet unknown constituents.  It could even be that there is some truly
emergent description of the electroweak phase
transition, a residual force that arises from the strong dynamics among
the weak gauge bosons~\cite{chano}.  We know that if we take the mass of the Higgs
boson to very large values, beyond a TeV in the Lagrangian of the
electroweak theory, the scattering among gauge bosons becomes strong,
in the sense that $\pi\pi$ scattering becomes strong on the GeV scale.
Resonances form among pairs of gauge bosons, multiple production of
gauge bosons becomes commonplace, and that resonant behavior could be
what hides the electroweak symmetry.  A new thought is that electroweak
symmetry breaking is the echo of extra spacetime 
dimensions~\cite{csaki}.  We don't
know, and we intend to find out during the next decade which path
nature has taken.

One very important step toward understanding the new force is to find
the Higgs boson and to learn its properties.  I've said before in
public, and I say again here, that the Higgs boson will be discovered
whether it exists or not.  The precise technical meaning of my
assurance is this.  There will be (almost surely) a spin-zero object
that has effectively more or less the interactions of the
standard-model Higgs boson, whether it is an elementary particle that
we put into to the theory or something that emerges from the theory.
Such an object is required to make the electroweak theory behave well
at high energies, once electroweak symmetry is hidden. You will find 
it, and that will be the start of something big~\cite{sparis}.

\section{Revolution: The Meaning of Identity}
The second revolutionary theme is one that I suspect will take much 
longer to define and achieve; it has to do with the tantalizing 
question of ``What makes a top quark a top quark, an electron an 
electron, and a neutrino a neutrino? What distinguishes these 
objects?'' In more operational terms, we may ask, ``What determines 
the masses and mixings of the quarks and leptons?'' It is not enough 
to answer, ``The Higgs mechanism,'' because the fermion masses are 
a very enigmatic element of the electroweak theory. Indeed, 
\textit{all fermion masses, starting with the electron mass, are 
evidence for physics beyond the standard model!} Once the electroweak 
symmetry is broken, our theory permits---welcomes---fermion masses, 
but the values of the masses are set by the famous, and apparently 
arbitary, Yukawa couplings of the Higgs boson to the fermions. Nothing 
in the electroweak theory is ever going to prescribe those couplings. 
It is not that the calculation is technically challenging; there is 
no calculation.

The exciting prospect, then, is that quark and lepton masses, mixing 
angles, and $\mathcal{CP}$-violating phases put us in contact with 
physics beyond the standard model. The challenge for us is to 
construct what the big question really is. We know very well what 
measurements we would like to make in $B$ physics, charm and strange 
physics, and neutrino physics---which elements of the mixing matrices 
we would like to fill in and which relationships we would like to 
test. Perhaps we will find that these don't fit the framework in 
which we view them, and that will give us some insight into the new 
physics. But if our standard-model framework passes every test, the 
new physics will still be there, and we need to understand how to get 
at it. There is a role here for the LHC, to be sure.

We may find new phenomena that suggest the origin of some or all of 
the quark and lepton masses.\footnote{Perhaps lepton flavor violation 
will emerge as an important clue.} And it might just be that we haven't 
grasped a latent pattern in the masses because we're not seeing the 
whole picture yet. Perhaps it will take discovering a new kind of 
matter---superpartners, or something entirely different---and seeing 
the spectrum of those new particles before it all begins to make 
sense. I do think it important that we consider the quarks and leptons 
together, to learn whether neutrino mass truly stands apart, and 
whether a common analysis can bring new insights.

I also believe that this question will, in the end, have 
revolutionary impact, once we understand what the question is. So it 
is up to all of us---not just to LHC$b$, not just to the 
flavor-physics groups in ATLAS, CMS, and ALICE---to pay attention to 
the problem of identity, and to learn how to frame the question. 
Lifting the veil of electroweak symmetry breaking will be a big step, 
but I cannot guarantee that it will suffice.

Until now, our best hope for finding simple relations among quark and 
lepton masses has come from unified theories of the strong, weak, and 
electromagnetic interactions. Those theories are the focus of our next 
topic.

\section{Revolution: The Unity of Quarks \& Leptons}
The quarks and leptons have many attributes in common.  All are
spin-$\cfrac{1}{2}$ particles, structureless at our current limits of
resolution, and the six quarks seem paired with the six leptons.  But
could we have a world made only of quarks, which respond to the strong
interaction, or only of leptons, which do not? If the known quarks 
and leptons were unrelated sets that matched by chance, how could we 
account for the remarkable neutrality (to 1 part in $10^{22}$) of 
ordinary matter? It seems unreasonable to us that the surpassing 
balance between the proton charge and the electron charge could be 
mere coincidence. Thus we are led to imagine quarks and leptons as 
members of an extended family, and from that hypothesis flows the full 
story of unified theories.\footnote{In less down-to-earth terms, we 
require matched pairs of quarks and leptons in order that the 
electroweak theory be free of anomalies, and so make sense up to high 
energies in the presence of quantum corrections.}

Once we assign color-triplet quarks and color-singlet leptons to the
same unified-theory multiplet, it is a natural implication that protons
should decay, mediated by quark--lepton transformations.  That natural
implication might not be unavoidable, because we don't know which
quarks go with which leptons. The traditional pairings: up and down 
with the electron and its neutrino, etc., are based only on 
tradition---the order in which we encountered the particles.  For 
all we know \textit{experimentally,} the first generation of quarks 
might go with the third generation of leptons. If we can find evidence 
for proton decay, we shall have definitive proof of the connection 
between quarks and leptons and gain information on which quarks should 
be associated with which leptons. Supersymmetric unified theories 
offer an attractive target of a proton lifetime only one or two 
orders of magnitude away from the current lower bound. Our challenge 
is to design the massive, low-background, finite-cost detector to push 
the sensitivity by a factor of 100.

A characteristic prediction of unified theories is coupling-constant 
unification, the statement that at some high-energy scale, the 
apparently independent couplings we observe in our low-energy world 
all have a common value. To test a candidate unified theory, we evolve 
the measured low-energy values up to high energies---which entails 
some assumptions about the spectrum of particles between here and 
there---and see whether the values coincide somewhere. That's a 
familiar, and valuable, exercise that has given us a hint of 
TeV-scale supersymmetry. But another way to view the same system of 
equations is to imagine that, on some happy day in the future, a 
theory might tell us the unification scale and the value there of the 
unified coupling constant. Then the differing values we see at low 
energy for the U(1) associated with weak hypercharge, the SU(2) 
associated with weak isospin, and the SU(3) associated with color 
come about because of the different evolution given by the different 
gauge groups and the particle spectrum. In this sense, we can 
understand why the strong interaction becomes strong on a certain 
scale.

There is a parallel to the running of coupling constants in the 
running of particle masses. Perhaps the pattern of quark and lepton 
masses looks weird---pattern\textit{less}---to us because we measure 
the masses at low scales, and not at the high scale where the values 
are set. At the appropriate high scale, the pattern might be 
rational---literally!---given by symmetry factors of some sort. 
According to our current spotty interpretation, the masses and mixing 
angles arise together, so any such exercise must confront plenty of 
constraints.

It seems to me that one of our urgent goals should be to understand 
how we can establish the unity between quarks and leptons, and how we 
would follow up the discovery of quark--lepton transitions.

\section{Revolution: Gravity Rejoins Particle Physics Rejoins Gravity \ldots}
For good reason, particle physicists normally neglect the influence of
gravity on particle collisions or decays.  If we estimate the rate for
a representative process, such as kaon decay into a pion plus a
graviton, it's easy to see that the emission of a graviton is
suppressed by $M_{K}/M_{\mathrm{Planck}}$.  The Planck mass
($M_{\mathrm{Planck}} \equiv (\hbar c/G_{\mathrm{Newton}})^{1/2}\approx
1.22 \times 10^{19}\gev$) is a big number because Newton's constant is
small in the appropriate units.  A dimensional estimate for the
branching fraction is $B(K \to \pi G) \approx
({M_{K}}/{M_{\mathrm{Planck}}})^{2} \approx 10^{-38}$.  It will be a
long time before the single-event sensitivity of any  experiment
reaches this level!

One realm in which gravity has not been far from our thoughts involves
the problem of separating the electroweak scale from higher scales.
The electroweak scale is not the only one we recognize as significant:
the Newtonian theory of gravity points to the Planck scale, and there
may be a unification scale for strong, weak, and electromagnetic
interactions; for all we know, there are intermediate scales, where
flavor properties are determined and masses are set.  The essence of
the hierarchy problem, as it is called, is this: We know that the
Higgs-boson mass must be less than a TeV, but the scalar mass
communicates quantum-mechanically with the other scales that may range
all the way up to $10^{19}\gev$.  How do we keep the Higgs-boson mass
from being polluted by the higher scales?

Our response, for twenty-five years or so, has been to seek to extend
the standard model, to temper the influence of distant mass scales.
Supersymmetry, which balances fermion loops against boson loops, is one
richly elaborated example~\cite{kraml,paige}, and the notion that the Higgs boson may be
composite is another.  Now a new approach is under investigation,
spurred by the recognition that we have investigated the electroweak
theory and the rest of the standard model up to about $1\tev$, but we
have  tested the inverse-square law of gravity only up to energies of
10~meV (yes, \textit{milli}-electron volts)! We have turned the question
around to ask why the Planck scale is so much bigger than the
electroweak scale, rather than why the electroweak scale is so low.  In
other words, why is gravity so weak? We'll see some consequences of 
posing the problem this way in the following Section.

Gravity has also weighed on the minds of electroweak insiders for three
decades.  In the electroweak theory, all of space is pervaded by a
vacuum energy density that turns out to be really large.  The
contribution of the Higgs field's vacuum expectation value to the
energy density of the universe is $\varrho_{H} \equiv
{M_{H}^{2}v^{2}}/{8}$, where $M_{H}$ is the Higgs-boson mass and $v
\approx 246\gev$ is the scale of electroweak symmetry breaking.  A
vacuum energy density corresponds to a cosmological constant $\Lambda =
{ ({8\pi G_{\mathrm{Newton}}}/{c^{4}}}){\varrho_{\mathsf{vac}}}$ in
Einstein's equations.  We've known for a very long time that there is
not much of a cosmological constant, that the vacuum energy has to less
than about $\varrho_{\mathrm{vac}} \ltap 10^{-46}\gev^{4}$, a very
little number.  It corresponds to $\approx 10\mev/\ell$ or
$10^{-29}\hbox{ g}\cm^{-3}$.  

But if we use the current lower limit on the Higgs-boson mass, $M_{H}
\gtap 114\gev $, to estimate the vacuum energy in the electroweak
theory, we find $\varrho_{H} \gtap 10^{8}\gev^{4}$.  That is wrong by
no less than fifty-four orders of magnitude!  This mismatch has given
many of us a chronic dull headache for about thirty years.  In the
simplest terms, the question is, ``Why is empty space so nearly
massless?''  A new wrinkle to the vacuum energy puzzle is the evidence
for a nonzero cosmological constant, respecting the bounds cited a
moment ago.  That discovery recasts the problem in two important ways.
First, instead of looking for a principle that would forbid a
cosmological constant, perhaps a symmetry principle that would set it
exactly to zero, now we have to explain a tiny cosmological constant!
Second, from the point of view of the dialogue among observation and
experiment and theory, now it looks as if we have access to some new
stuff whose properties we can measure.  Maybe that will give us the
clue that we need to solve this old problem.

\section{Revolution: A New Conception of Spacetime}
Asking why gravity is so weak has given rise to new thinking, part of it
connected with a new conception of spacetime.  What is our evidence
that spacetime is really three-plus-one dimensional?  How well do we
know that there are not other, extra, dimensions?  What must be the
character of those extra dimensions, and the character of our ability
to investigate them, for them to have escaped our notice?  How can we
attack the question of extra dimensions experimentally?

I will just call attention to a few examples of how physics might be 
changed if additional dimensions have eluded detection.

Perhaps, in contrast to the strong and electroweak gauge forces,
gravity can propagate in all dimensions, including those we haven't
perceived, because it is universal.  When we inspect the world on small
enough scales, we will see gravity leaking into the extra dimensions.
Then by Gauss's law, the gravitational force will not be an
inverse-square law, but will be proportional to $1/r^{2+n}$, where $n$
is the number of extra dimensions.  That would mean that, as we
extrapolate to smaller distances, or higher energies, gravity will not
follow the Newtonian form forever, as we conventionally suppose.  On
small scales, gravity will evolve more rapidly; its strength will grow
faster, and so it might rejoin the other forces at a much lower energy
than the Planck scale we have traditionally assumed.  That could change
our perception of the hierarchy problem entirely.

Perhaps extra dimensions offer a new way to try to understand the
dramatic hierarchy of fermion masses, ranging from $1$ for the top
quark, in natural units of the Higgs field's vacuum expectation value,
to a few $\times 10^{-6}$ for the electron, and so on. 
If gravity is intrinsically strong but spread out into many 
dimensions, tiny black holes might be formed in high-energy 
collisions, and we might just be able to detect the exchange or 
emission of Ka\l uza--Klein towers of gravitons at the 
LHC~\cite{ljoe,smaria}. While gravity is generally negligible in 
particle-physics processes at low energies, it has been present in 
our consciousness for years in the vacuum energy problem and the 
hierarchy problem of the electroweak theory. Perhaps now gravity is 
presenting itself as an opportunity---one that is here to stay!

\section{Envisioning Particles and Interactions}
I have been concerned for some time with the prevailing narrow view of
the goals of our science.  It distresses me to read in the popular
press that the sole purpose of the LHC is to find---to check off, if
you will---the Higgs boson, the holy grail (at least for this month) of
particle physics.  I am troubled still more when the shorthand of the
Higgs search narrows the discourse within our own community.  In
response, I have begun to evolve a visual metaphor---the double
simplex---for what we know, for what we hope might be true, and for the
open questions raised by our current understanding.  I have a deep
respect for mathematics as a refiner's fire, but I believe that we should
be able to explain the essence of our ideas in languages other than
equations.  I interpolated a brief animated overview~\cite{EnvPIQT} of
the double simplex at this point in my lecture.  For a preliminary
exposition in a pedagogical setting, see Ref.~\cite{Quigg:2004is}.  I
am at work on more complete explanation of the aims of particle physics
through the metaphor of the double simplex.

\section{Observations, Opportunities, Concerns}
Before concluding, I would like to offer some remarks in the spirit of Jos Engelen's 
message~\cite{ejos} to ``friendly laboratories.''

\textit{To CERN:} Keep your focus on the LHC to make it a glorious
success---\textit{soon!} You carry the hopes and dreams of us all.

\textit{To CERN and ECFA and EPS:} Please find ways to welcome others
into Europlanning.  The Summer Study on the Future of Particle Physics 
held at Snowmass in 2001 was immeasurably strengthened and enriched by 
the participation of more than two hundred colleagues from outside the 
United States. We need to do more, in all regions, to draw 
enlightenment and support---and even hard questions---from our 
colleagues around the world.

\textit{To all the rulers of the particle physics universe:} We thrive
on competition, but hyperunilateralism will be our common undoing.  We
all have a stake in a healthy \textit{world} program.

\textit{To the LHC community:} Respect the Tevatron, hope for some
vigorous competition, and learn from the Tevatron experience~\cite{tev4lhc}.

\textit{To all of us:} We have an obligation to involve more people in
the adventure of our science, our trust in experiment over authority,
and our shared belief in the power of reason and the importance of
doubt.  We celebrate the many nations represented in the LHC
collaborations, but I draw your attention to the vast blank expanses on the
CMS and ATLAS collaboration maps. Those parts of the world need our 
ideas and our technical expertise, too; and we need to engage the 
people of those parts in the scientific enterprise.

\textit{To the LHC community: How will we actually do physics at the
LHC?} Can everyone who wants  to participate be accommodated at
CERN? What would be required for people to participate effectively at
regional analysis centers?  (Of the centers?  of CERN? of the
experiments?)

\textit{To all of us:} How can we advance the commissioning of the
Right Linear Collider?  Must we execute projects in sequence?  Can we
optimize scientific return by executing in parallel, through
cooperation and global networks?  How can we best take advantage of the
multitude of the scientifc opportunities before us~\cite{balain}?

\section{The Road Ahead}
The Large Hadron Collider will lead us into real golden age of
exploration and discovery.  I look forward to a wonderful flowering of
experimental particle physics, and of theory that engages with
experiment.  \P\ We will make a thorough exploration of the 1-TeV
energy scale; search for, find, and study the Higgs boson or its
equivalent; and probe the mechanism that hides electroweak symmetry.
\P\ We will continue to challenge the standard model's attribution of
$\mathcal{CP}$ violation to a phase in the quark mixing matrix, in
experiments that examine $B$ decays and rare decays---or mixing---of
strange and charmed particles.  \P\ New accelerator-generated neutrino
beams, together with reactor experiments and the continued study of
neutrinos from natural sources, will consolidate our understanding of
neutrino mixing.  Double-beta-decay searches may confirm the Majorana
nature of neutrinos.  \P\ The top quark will become an important window
into the nature of electroweak symmetry breaking, rather than a mere
object of experimental desire.  Single-top production and the top
quark's coupling to the Higgs sector will be informative.  \P\ The
study of new phases of matter, especially through heavy-ion collisions,
and renewed attention to hadronic physics will deepen our appreciation
for the richness of QCD, and might even bring new ideas to the realm of
electroweak symmetry breaking.  \P\ Planned discoveries and
programmatic surveys have their (important!)  place, but exploration
breaks the mold of established ideas and can recast our list of urgent
questions overnight.  Among the objectives we have already prepared in
great theoretical detail are extra dimensions, new strong dynamics,
supersymmetry, and new forces and constituents.  Any one of these would
give us a new continent to explore.  \P\ Proton decay remains the most
promising path to establish the existence of extended families that
contain both quarks and leptons.  A vast new underground detector will
be required to push the sensitivity frontier.  \P\ We will learn much
more about the composition of the universe, perhaps establishing the
nature of some of the dark matter.  Observations of type Ia supernovae,
the cosmic microwave background, and the large-scale structure of the
universe will extend our knowledge of the fossil record.  Underground
searches may give evidence of relic dark matter.  Collider experiments
will establish the character of dark-matter candidates and will make
possible a more enlightened reading of the fossil record.

You who have been preparing the LHC experiments know that none of this 
will be easy. We have miles to go before the beams cross, the 
detectors record events, and we begin to decipher the messages they 
contain. At the same time, we are taking such a great step into the 
unknown---including the \onetev\  where we know many treasures are 
hidden---that I believe the flood of amazing results will come 
quickly, while we are still learning how to listen to our 
detectors~\cite{fab}. It is a glorious prospect; how lucky we are to 
be part of it!
    
\bigskip

{\small Fermilab is operated by
Universities Research Association Inc.\ under Contract No.\
DE-AC02-76CH03000 with the U.S.\ Department of Energy.  I commend the 
members of the organizing committee and their staff for the 
impeccable welcome in Vienna and for the well-chosen program, both 
scientific and cultural. And thanks to the speakers and other 
participants for the high quality of the talks and the discussions, 
both formal and informal. I look forward to LHC2006 in Krakow.

\end{document}